\documentclass[aps,pre,twocolumn]{revtex4}
\input epsf
\usepackage{graphicx}
\usepackage{dcolumn}
\usepackage{amsmath}
\usepackage{bm}
\usepackage{color}

\begin{document}
\title{Dewetting of solid films with substrate mediated evaporation
}
\author{Anna Chame$^1$, Olivier Pierre-Louis$^2$, }
\affiliation{$^1$ Instituto de F\'{\i}sica, Universidade Federal Fluminense,
 24210-340 Niter\'oi RJ, Brazil\\
$^2$ LPMCN, Université Lyon 1, 43 Boulevard du 11 Novembre 1918
F 69622 Villeurbanne, France.
}
\date{\today}

\begin{abstract}
The dewetting dynamics of an ultrathin film is studied in the presence of
evaporation --or reaction-- of adatoms on the substrate. KMC simulations
are in good agreement with an analytical model with diffusion, rim facetting, and substrate sublimation.
As sublimation is increased, we find a transition from the usual dewetting
regime where the front slows down with time, to a sublimation-controlled regime where
the front velocity is approximately constant. The rim width exhibits an unexpected
non-monotonous behavior, with a maximum in time.
\end{abstract}

\maketitle
%
\section{Introduction}

Dewetting, the process by which a thin film breaks up into
droplets, is a non-equilibrium
process driven by the reduction of the total energy.
A large number of recent experimental studies
have been devoted to the analysis of the dewetting of
ultra-thin solid 
films\cite{Jiran,Yang2005,barbe,Yang2005,Danielson2006,Krause,Luber2010,Galinski2010,Bussmann2011}. 
The main mass transport process during dewetting
is surface diffusion. However, early work  from Srolovitz and Safran\cite{Srolovitz1986}
on solid dewetting have already pointed out the role of 
sublimation, which cannot be avoided at high temperatures. 
These authors have discussed the case of sublimation
directly from the film to the vapor.
Here, we discuss the regime where atoms from the 
film diffuse on the substrate, and may evaporate
when they are on the substrate.

Further motivation of the present study comes from 
recent work on the dewetting of SOI systems (Si/SiO$_2$), which
have revealed the presence of reaction of Si on the substrate
(with subsequent sublimation of the products formed in the reaction)
\cite{Danielson2006,Sudoh2010}.
Such a phenomenon could be modeled by a simple sublimation
rate to a first approximation, neglecting the substrate
shape changes induced by the etching related to the reaction.
In addition, sublimation could also be relevant for the
dewetting of metallic films where the temperature can be
raised close to the melting temperature\cite{Luber2010,Galinski2010}.

In this article, we analyze the detailed dynamics
of dewetting with the possibility for atoms to leave the film,
diffuse on the substrate, and evaporate.
We perform Kinetic Monte Carlo (KMC) simulations,
and provide a model inspired from that of Ref.\cite{PierreLouis2009},
which is in good agreement with 
KMC simulation results.
We show that substrate evaporation leads to a novel dynamical regime
where the film edge velocity is approximately constant.
In addition, the rim width exhibits an unexpected non-monotonous 
behavior in time, reaching a maximum value at some intermediate stage.

\section{KMC simulations}

We model the dewetting of a crystalline film using a solid-on-solid model
on a 2D square lattice. The substrate is flat and frozen, and is associated
to height $z= 0$, whereas the film has local height $z \ge 1$. The local height
will change as atoms diffuse or evaporate. There is no incoming flux of
atoms in this model.

An atom at the surface can hop to nearest neighbor
sites with rates $\nu_n$, when it is not in direct contact with the substrate,
and $r_n$ when it is in contact with the substrate ($z= 1$).
In our model an atom has to break all its bonds to
hop. The hopping barrier is therefore given by the binding energy of the atom.
An isolated atom  over the substrate can also evaporate with rate $r_e$.
Then:
\begin{eqnarray}
\nu_n &=& \nu \,{\rm e}^{-(nJ+J_0)/T },
\label{e:nu_n}
\\ 
r_n &=& \nu \,{\rm e}^{-(nJ+J_0-E_S)/T},
\label{e:r_n}
\\
r_e &=& \nu \,{\rm e}^{-(J_0+E_{vs}-E_S)/T },
\label{e:r_e}
\end{eqnarray}
where $\nu$ is an attempt frequency, $T$ is the temperature
(in units with $k_B = 1$), $n$ is the number of in-plane nearest neighbors
of the atom before the hop, $J$ is the lateral bond energy between two atoms in
the film, $J_0$ is the vertical bond energy between two atoms in the film,  
$E_{vs}$ is the energy barrier for desorption of an atom in contact
with the substrate and $E_S$ is the adsorbate-substrate  excess energy.
The model is presented in Fig1. We choose J as the energy unit, so that
$J= 1$ in the following. Since $J_0$ appears in all rates in 
Eqs.(\ref{e:nu_n},\ref{e:r_n},\ref{e:r_e}), we choose $\nu_0^{-1}$ as the time unit, where
\begin{eqnarray}
\nu_0=\nu\, {\rm e}^{-J_0/T}.
\end{eqnarray}
Hence, in the following, we set $\nu_0=1$.

\begin{figure}
\includegraphics[height=4 cm]{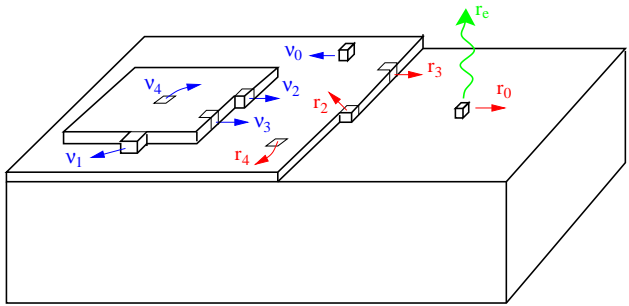}\\
\caption{(Color online) Schematics of the Kinetic Monte Carlo (KMC) model, with hopping rates
$\nu_n$ and $r_n$, and sublimation rate $r_e$.
}
\label{fig1}
\end{figure}

The parameter $E_S$ controls the wetting properties of the film on the
substrate. For example, when $E_S \le 0 $, the film completely wets the substrate.
In this work, we  consider  $ E_S  > 0 $ (partial wetting).
It can be shown \cite{PierreLouis2010} that the surface excess energy $E_S$ can be
written as
\begin{eqnarray}
E_S=E_{AV}+E_{AS}-E_{SV},
\end{eqnarray}
where $E_{AV}$, $E_{AS}$ and $E_{SV}$ are  the energy per site of the 
adsorbate-vacuum surface, adsorbate-substrate interface 
and substrate-vacuum surface, respectively.
This parameter is then identical (with opposite sign) to the
spreading coefficient S defined in references
Ref.\cite{deGennes1985,Sangiorgi1988} for the study of liquid dewetting.

The algorithm used in the simulations is the following.
We  list all atoms into  classes. Each class
is characterized by the number of  in-plane  neighbors $n$ of the atom
and by the existence or not of  a nearest neighbor  belonging to the
substrate. At a given time t, we calculate the probabilities per unit
time $w_i$ of all possible events (an event is the motion  of an atom
originally at position $i$) given either by Eq.(\ref{e:nu_n}),
Eq.(\ref{e:r_n}) or Eq.(\ref{e:r_e}), and the sum W of all those rates
(for all mobile atoms). We increment the time by a $\delta t$, which is
equal to the inverse of the sum of the rates of all possible events
\footnote{This choice for $\delta t$  corresponds to  the average value of the waiting
time between two successive events \cite{Kotrla1996}.}, $1/W$.
We choose the event with probability $w_i/W$. In the case of diffusion, the atom
moves with equal probability in any of the four possible  directions.

\begin{figure}
\includegraphics[height=12 cm]{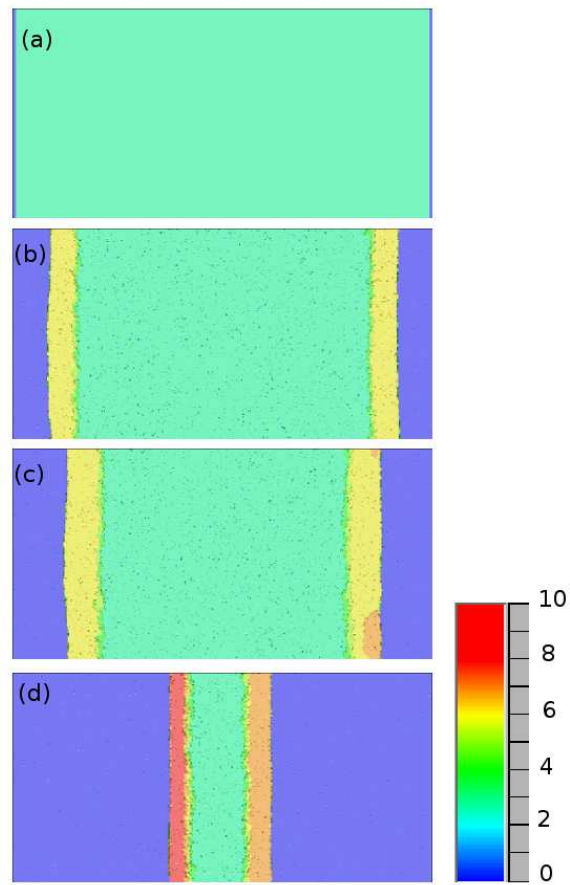}\\
\caption{(Color online) Snapshots of the dewetting dynamics with substrate evaporation observed in KMC.
(a) We start from an initial trench in the film. (b,c,d) Two dewetting rims form
and move in opposite directions. Note that the rim width exhibits
a maximum in the intermediate stages in (c).
Simulation parameters: $E_{vs}=5$, $L_x= 400$, $L_y=800$, $T=0.4$, $E_S= 0.5$ and $h=3$.
The $z$ height scale of the surface is indicated on the right.
}
\label{fig2}
\end{figure}

The initial configuration
is a film of constant height covering  the substrate, leaving only
a small stripe uncovered along the (10) direction. The film edges are straight in
the beginning. Atoms diffuse at the surface of the film and on
the substrate. However, they can evaporate only when they are on the substrate.
If we let the system evolve for a long enough time, all atoms should
evaporate, and there will be no atoms left on the substrate in the final state.
However, we are only interested in the early stages of the 
dynamics, where the dewetting dynamics occurs. As shown in Fig.\ref{fig2},
the dewetting dynamics proceeds with the formation of receding dewetting rims
at the edges of the film.

We varied  the sublimation energy barrier $E_{vs}$ to investigate  the
influence of sublimation in the dewetting regimes. If $E_{vs}/T$ is
sufficiently high, we recover the case considered before
\cite{PierreLouis2009,PierreLouis2010}
with no sublimation (the adsorbate dewets  driven only
by diffusion and attachment/detachment of adatoms).

\section{Model}
\label{s:model}
\subsection{Rim width and position dynamics}

In order to analyze the dewetting dynamics observed in KMC simulations,
we use a 1D model in which the dewetting front is assumed to be straight,
and invariant along the direction $y$.
The position of the edge of the film is denoted as $x_1$.
At $x_1$, the height exhibits a jump of height $h_1$ between
the substrate and the rim top facet. On the other side
of the dewetting rim, a bunch of atomic steps is at position $x_2>x_1$.
On the film for $x>x_2$, the height stays at the value of the initial film height $h$.
Defining $h_2=h_1-h$, we have:
\begin{eqnarray}
h_1\partial_tx_1&=&-\Omega  D\partial_xc|_1+(x_2-x_1) \partial_th_1-\Omega D_s\partial_x c_s|_1,
\label{e:flat_1}
\\
h_2\partial_tx_2&=&-\Omega D\partial_xc|_2,
\label{e:flat_2}
\end{eqnarray} 
where $c$ and $D$ are the adatom concentration
and  diffusion constant on the rim top facet, and
$c_s$ and $D_s$ are the adatom concentration
and  diffusion constant on the substrate.
In addition, $\Omega$ is the atomic volume.

On the substrate, the adatom concentration $c_s$ obeys a quasistatic
diffusion equation with evaporation
\begin{eqnarray}
D_s\Delta c_s-{c_s\over \tau_s}=0,
\label{e:diff_substrate}
\end{eqnarray}
where $D_s$ and $\tau_s$ are the diffusion constant
and sublimation time of adatoms on the substrate.
The general solution of Eq.(\ref{e:diff_substrate}) reads:
\begin{eqnarray}
c_s=A\cosh (x/x_s)+B\sinh(x/s_s),
\label{e:cs_of AB}
\end{eqnarray}
where $x_s=(D_s\tau_s)^{1/2}$ is the sublimation length.
We consider a geometry similar to that of KMC simulations:
Two dewetting fronts, separated by the distance $\ell_s=2x_1$,
recede in opposite directions.
The boundary condition at the fronts is assumed to be
instantaneous attachment-detachment kinetics, leading to:
\begin{eqnarray}
c_s|_{1}=c_{eq}^s.
\end{eqnarray}
Using this condition in Eq.(\ref{e:cs_of AB}), we find
\begin{eqnarray}
c_s=c_{eq}^s{\cosh (x/x_s)\over \cosh (\ell_s/2x_s)},
\end{eqnarray}
and, as a consequence the contribution of evaporation to the dewetting front
velocity in Eq.(\ref{e:flat_1}) is:
\begin{eqnarray}
v_s\equiv \Omega D_s\partial_xc_s|_1=\Omega \left(D_s \over \tau_s\right)^{1/2}c_{eq}^s\tanh(x_1/x_s).
\end{eqnarray}

On the top of the rim, the adatom concentration $c$
obeys a quasistatic diffusion equation
\begin{eqnarray}
\Delta c=0.
\end{eqnarray}
Assuming translational invariance in the other direction $y$,
and instantaneous attachment-detachment
kinetics leading to fixed concentrations $c_1=c_{eq}\exp(E_S/h_1T)$ and $c_2=c_{eq}$ at $x_1$ and $x_2$,
we obtain for $x_1\leq x\leq x_2$:
\begin{eqnarray}
c=c_1 +{x-x_1\over x_2-x_1}(c_2-c_1),
\label{e:linear_conc_profile}
\end{eqnarray}
so that $\partial_xc|_1=\partial_xc|_2=(c_2-c_1)/(x_2-x_1)$.
The dynamical equations may then be re-written as:
\begin{eqnarray}
\partial_tx_1&=&-{\Omega D\over h_1}\,{c_2-c_1\over x_2-x_1}
+(x_2-x_1) {\partial_th_1 \over h_1}-{v_s\over h_1},
\label{e:dynamical_flat1}
 \\
\partial_tx_2&=&-{\Omega D\over h_2}\,{c_2-c_1\over x_2-x_1}.
\label{e:dynamical_flat2}
\end{eqnarray}
Now $h_1$ and $h_2=h_1-h$ are time dependent. Moreover,
$c_2-c_1$ also depends on time via $h_1$.

\subsection{Rim height evolution}

Since we assume no evaporation on the film, we use the same theory
as in Ref.\cite{PierreLouis2009} for the rim height evolution. 
Following Ref.\cite{PierreLouis2009} we assume that the rim height evolution
is controlled by 2D nucleation on the facet with a nucleation rate:
\begin{eqnarray}
{\cal J}=\ell\,{Dc_{eq} \over \Omega}
\left(E_S \over Th_1\right)^{3/2}\left(T^2\over \pi\gamma^2\Omega\right)
\,{\rm e}^{-\pi \gamma^2 \Omega h_1 /TE_S},
\label{e:nucleation_top_facet}
\end{eqnarray}
where $\gamma$ is the atomic step free energy on the film.
The monolayer islands formed by the 2D nucleation process then spread
on the rim top facet with the velocity\cite{PierreLouis2009}:
\begin{eqnarray}
V_{zip}\approx  C_{zip}\Omega D c_{eq}  {E_S^2 \over \Omega T h_1^2\tilde \gamma},
\label{e:Vzip}
\end{eqnarray}
where $C_{zip}\approx 0.25$ is a number. We will also assume that the temperature is
high enough so that steps are isotropic, and $\tilde\gamma\approx\gamma$.

Assuming a total rim length $L$, the rim height evolution reads\cite{PierreLouis2009}:
\begin{eqnarray}
\partial_th_1={\cal J} \min\left[L,\left(V_{zip}\over {\cal J}\right)^{1/2}\right],
\label{e:h1_evol_MNSN}
\end{eqnarray}
which accounts for both regimes of single and multiple nucleation.

The model equations (\ref{e:dynamical_flat1},\ref{e:dynamical_flat2},\ref{e:h1_evol_MNSN})
are similar to that of Ref.\cite{PierreLouis2009}. The only difference is the
term $v_s$ which appears in Eq.(\ref{e:dynamical_flat1}).

\section{Comparison between model and KMC}

\subsection{KMC model parameters}
In this section, we shall evaluate the model parameters as a function
of the KMC parameters.

On the film, the adatom concentration is given by detailed balance\cite{PierreLouis2009}:
\begin{eqnarray}
c_{eq}=\Omega^{-1}{\rm e}^{-2J/T},
\end{eqnarray}
and the diffusion constant reads:
\begin{eqnarray}
D = {\Omega \over 4}.
\end{eqnarray}
The energy cost for an adatom to be in contact with the substrate
is $E_S$, and therefore:
\begin{eqnarray}
c_{eq}^s&=&\Omega^{-1}{\rm e}^{(-2J-E_S)/T},
\\
D_s &=& {\Omega \over 4}{\rm e}^{E_S/T}.
\end{eqnarray}
Finally, the sublimation time on the substrate reads:
\begin{eqnarray}
\tau_s&=&{\rm e}^{(E_{vs}-E_S)/T},
\end{eqnarray}
where $E_{vs}$ is the substrate desorption energy.

\subsection{Quantitative comparison}
We have measured $x_1$, $h_1$, and $\ell$
as a function of time for various parameters in KMC simulations.
As opposed to the case without sublimation,
we have not found any (even partial) analytical solution
of the model. We therefore resort to
a full numerical solution of
Eqs.(\ref{e:dynamical_flat1},\ref{e:dynamical_flat2},\ref{e:h1_evol_MNSN})

\begin{figure}
\includegraphics[height=6 cm]{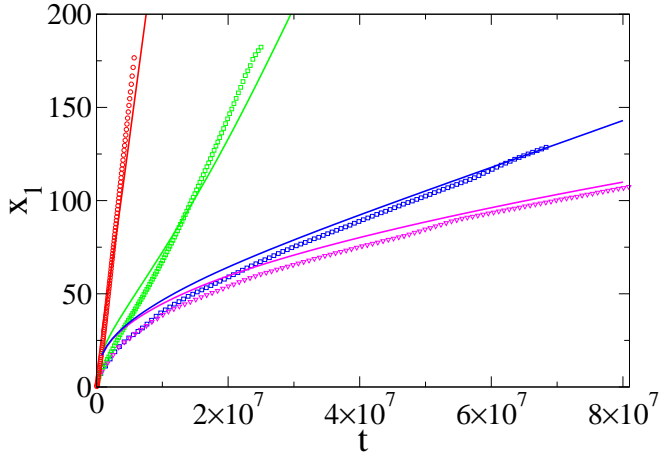}\\
\caption{(Color online) Front position $x_1$ for $E_{vs}=3$ (red circles), $4$ (green squares),
$5$ (blue squares),and $6$ (magenta triangles).
Other parameters: $E_s=0.5$, $L_x=400$, $L_y=400$, $T=0.4$, and $h=3$.
Symbols are results from KMC simulations, and the solid lines
represent the numerical solution of the model of section \ref{s:model}.
}
\label{fig3}
\end{figure}

\begin{figure}
\includegraphics[height=6 cm]{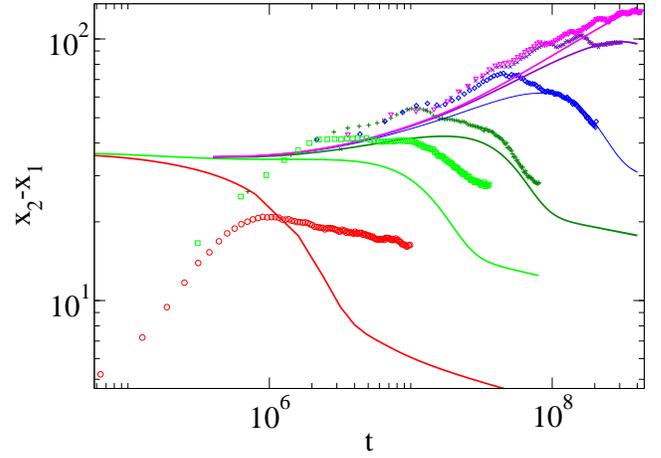}\\
\caption{(Color online) Rim width for $E_{vs}=3$ (red circles), $4$ (green squares),
$4.5$ (dark green crosses), $5$ (blue diamonds), $5.5$ (purple x's), 
and $6$ (magenta triangles).
Other parameters: $E_s=0.5$, $L_x=800$, $L_y=400$, $T=0.4$, and $h=3$.
Symbols are results from KMC simulations, and the solid lines
represent the numerical solution of the model of section \ref{s:model}.
}
\label{fig4}
\end{figure}

\begin{figure}
\includegraphics[height=6 cm]{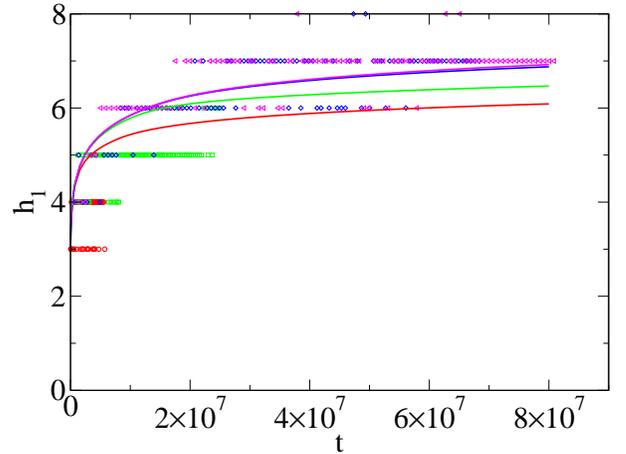}\\
\caption{(Color online) Rim height for $E_{vs}=3$ (red circles), $4$ (green squares),
$5$ (blue diamonds),and $6$ (magenta triangles).
Other parameters: $E_s=0.5$, $L_x=400$, $L_y=400$, $T=0.4$, and $h=3$.
Symbols are results from KMC simulations, and the solid lines
represent the numerical solution of the model of section \ref{s:model}.
}
\label{fig5}
\end{figure}

The agreement between KMC and the numerical solution of the
model was checked for $x_1$, $\ell$, and $h_1$. As shown 
in Fig.\ref{fig3},\ref{fig4},\ref{fig5}, the agreement
is good (note that there is no fitting parameter in the model).
Interestingly, the evolution of the rim width is non-monotonous, and exhibits a maximum
in time both in the model and in KMC simulations. 
However, the early-time dynamics is difficult
to reproduce since the first layer nucleation is not described 
correctly within our model, as seen in Fig.\ref{fig4}.
In addition, the measurement of the rim width in the early
stages of the dewetting process in KMC simulations is delicate. 
Indeed, we simply measure the area
of the film with height $z\geq h+1$, and divide by the total rim length $2L_y$.
Since adatoms and small monolayer islands on the film have $z=h+1$, they
are also counted. Hence, this procedure is inaccurate in the
initial stages when the total rim area is small.

\section{Discussion}

We have studied the dewetting dynamics of thin solid films
in the presence of substrate sublimation.
The overall agreement between KMC and the analytical model indicates that
we have caught the main physical ingredients of the dynamics.

\subsection{Transition time}

In this section, we discuss the transition between the 
early time diffusion-limited regime, and the late time sublimation-limited
regime.

Since the nucleation rate ${\cal J}$ decreases exponentially with $h_1$,
the evolution of $h_1$ essentially
occurs in a short transient regime at short times.
We shall therefore base our analysis on the fact that
$h_1$ evolves very slowly.

In the competition between diffusion and sublimation,
the process which controls the dynamics is the one which
leads to the largest contribution to the front velocity.
For fixed $h_1$, diffusion leads to $x_1\sim t^{1/2}$,
and sublimation leads to $x_1\sim t$. Hence, we expect that
diffusion always wins at short times, while sublimation controls the late
time regime.

Let us first consider the diffusion limited regime (where $v_s$ is 
negligible), with fixed $h_1$, leading to
\begin{eqnarray}
\ell_{Diff}= {h \over h_1-h} \left[2D\Omega (c_1-c_2) \left(h^{-1}-h_1^{-1}\right)t\right]^{1/2}.
\label{e:ell_diff}
\end{eqnarray}

\begin{figure}
\includegraphics[height=5 cm]{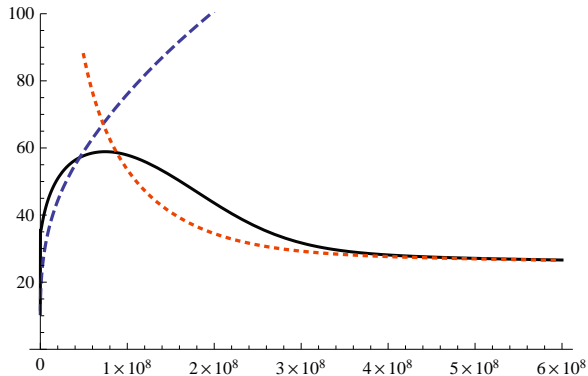}\\
\caption{(Color online) Rim width given by the model (solid line) for $E_{vs}=5$, $E_s=0.5$, $L_x=L_y=400$, $T=0.4$, and $h=3$.
The dashed line corresponds to $\ell_{Diff}$ Eq.(\ref{e:ell_diff}), and the dotted line to 
$\ell_{Subl}$ Eq.(\ref{e:ell_subl}).
}
\label{fig6}
\end{figure}

In the late-time dynamics, which is dominated
by sublimation, the front moves approximately at constant 
velocity. Hence, we expect the dewetting rim to be roughly
stationary, with a slow variation of $\ell$.
An evolution equation for $\ell$ is obtained by taking
the difference between Eq.(\ref{e:dynamical_flat1}) and Eq.(\ref{e:dynamical_flat2}),
and assuming that $\partial_th_1\ll v_s/\ell$.
Plugging $\partial_t\ell=0$ into this equation leads to
\begin{eqnarray}
\ell_{Subl}=D\Omega {c_1-c_2\over v_s} {h \over h_1-h}.
\label{e:ell_subl}
\end{eqnarray}
Recalling that $h_1$ is not fixed, but in fact increases slowly, 
the expression of $\ell_{Subl}$ should be interpreted as a
slow decrease of $\ell$ caused by the increase of $h_1$.
As shown on Fig.\ref{fig4}, $\ell_{Diff}$ and  $\ell_{Subl}$
provide reasonably good approximations of $\ell$ at short and
long times.

Combining the two expressions, we obtain the crossover time
from the relation $\ell_{Diff}=\ell_{Subl}$ as:
\begin{eqnarray}
t= {D\Omega c_{eq} \over 2v_s^2}\,{E_S \over T} {h \over h_1-h},
\end{eqnarray}
where we have assumed that $E_S\ll Th_1$. In addition, since the growth of
$h_1$ slows down exponentially with time, we assume that $h/(h_1-h)$ is of order
of 1. Finally, we obtain
\begin{eqnarray}
t\sim {D\Omega c_{eq} \over 2v_s^2}\,{E_S \over T}.
\end{eqnarray}
As a consequence, we see that the crossover time will decrease
as sublimation (and hence $v_s$) increases.
This formula provides the correct order of magnitude as compared
to the maximum observed in Fig.\ref{fig4}.

\subsection{Sublimation from the film}
The situation discussed in the present work
can be compared with the results of Srolovitz and Safran\cite{Srolovitz1986},
which also find a front evolution with constant velocity.
In their model, there is simultaneously sublimation
and adsorption between the film and the vapor, so that the net growth of
a flat film vanishes.
The motion of the film height $h$ is then driven
by local chemical potential variations. Since the local
chemical potential is proportional to the surface curvature,
we obtain in the small slope approximation: $\partial_th=A\partial_{xx}h$.
Typically, we expect $A=k\rho_{eq}\Omega\gamma_S/k_BT$,
where $k$ is the evaporation-condensation kinetic coefficient,
$\rho_{eq}$ is the equilibrium vapor density,
and $\gamma_S$ is the surface stiffness (which is assumed to be isotropic).

The edge of the film, which exhibits
a larger curvature, has a larger chemical potential, so that
the sublimation at the edge is reinforced and is not compensated
by adsorption from the vapor. Therefore, the front
recedes by sublimation at the film edge.
Srolovitz and Safran find a front velocity $V=A\tan\theta/\bar h$,
where $\bar h$ is initial film height, and $\theta$ is the contact angle,
which is related to $E_S$ via\cite{PierreLouis2009b}:
\begin{eqnarray}
E_S/J={1\over 2}(1-\cos(\theta)).
\end{eqnarray}
As a summary, the modeling of Srolovitz and Safran correspond
to the case of a film in equilibrium with its vapor, a situation which is very
different from the one considered in the present study. However,
the dewetting process qualitatively obeys the same behavior,
with fronts moving at constant velocity\footnote{Strictly speaking,
the front velocity is never constant in our model. Indeed,
a constant velocity requires a constant rim height $h_1$. Although
$\partial_t h_1$ decreases exponentially with $h_1$, it does not vanish,
so that $h_1$ is not constant.}.

\subsection{Discussion of experiments}
Dewetting of SOI systems, consisting of Si crystalline films
on amorphous SiO$_2$ substrates, has been investigated recently\cite{Yang2005,barbe,Bussmann2011}.
The presence of reaction Si+SiO$_2\rightarrow$2SiO has been suggested
in several experiments, and there is still an issue about the
determination of the place where the reaction occurs. If Si atoms diffuse
through the triple line on SiO$_2$, and then react with SiO$_2$,
then the situation would be quite similar to our model system.
(Note that the reaction should lead to an etching of the substrate,
which is not taken into account in our model). This scenario is consistent
with that proposed for the deoxydization of SiO$_2$ films on Si substrates
\cite{Miyata2000,Hibino2006}. However, the 
reaction mechanism is controversial, and a more complex scenario
involving the diffusion of oxygen at the
interface was also proposed\cite{Sudoh2010}. 
Therefore, we expect substrate reaction to lead to a change in the 
time evolution of the position of straight fronts,  from a power law 
behavior with an effective exponent close to $1/2$, to an approximately linear 
behavior. This crossover
could appear with increasing temperature if reaction is activated
at high temperatures. 

Substrate evaporation could also be important for the dewetting
dynamics of metallic films\cite{Luber2010,Galinski2010}, where the temperature is often
increased up to temperatures where sublimation cannot be neglected.

\section{Conclusion}

In conclusion, in the presence of adatom sublimation on the substrate, the motion
of a dewetting front exhibits a crossover from a constant thickening
and widening of the rim, accompanied with a slowing down (with
the front position obeying an approximate power-law $t^{1/2}$), to a regime where the
velocity is approximately constant.  In this latter regime,
the rim height increases slowly, and the rim width decreases slowly.
Surprisingly, the rim width exhibits a maximum at the crossover
between the two regimes. We have obtained quantitative agreement between
KMC simulations and a model with diffusion limited mass transport on the rim,
nucleation limited increase of the rim height, and evaporation on the substrate.

\end{document}